\documentclass[preprint,review,times,3p]{elsarticle}
\usepackage{graphicx}
\usepackage{subfig}
\usepackage[countmax]{subfloat}
\usepackage{amsmath, amssymb}

\biboptions{sort&compress,comma}

\journal{ Physica E}

\begin{document}

\begin{frontmatter}
\title{Edge electrostatics revisited}
\author[label1]{A. Salman}
\author[label1]{M. B. Yucel\corref{label5}}
\ead{yucelmb@gmail.com} \cortext[label5]{Corresponding
Author.
Tel.: +902423102278; Fax: +902422278911} 
\author[label2,label3]{A. Siddiki}
\address[label1]{Akdeniz University, Faculty of Sciences, Physics Department, Antalya 07058, Turkey}
\address[label2]{Istanbul University, Faculty of Sciences, Physics Department, Vezneciler-Istanbul 34134, Turkey}
\address[label3]{Harvard University, Physics Department, Cambridge 02138 MA, USA}

\begin{abstract}
In this work we investigate in detail, the different regimes of the
pioneering work of Chklovskii \emph{et al} \cite{Chklovskii92:4026},
which provides an analytical description to model the electrostatics
at the edges of a two-dimensional electron gas. We take into account
full electrostatics and calculate the charge distribution by solving
the 3D Poisson equation self-consistently. The Chklovskii formalism is
reintroduced and is employed to determine the widths of the
incompressible edge-states also considering the spin degree of
freedom. It is shown that, the odd integer filling fractions cannot
exist for large magnetic field intervals if many-body effects are
neglected. We explicitly show that, the incompressible strips which
are narrower than the quantum mechanical length scales vanish. We numerically and analytically show that, the
non-self-consistent picture becomes inadequate considering realistic Hall bar
geometries, predicting large incompressible strips. The
details of this picture is investigated considering device
properties together with the many-body and the disorder effects.
Moreover, we provide semi-empirical formulas to estimate realistic
density distributions for different physical boundary conditions.
\end{abstract}

\end{frontmatter}

\section{Introduction}
\label{intro} Cooling a high-mobility two-dimensional electron system
(2DES) near absolute zero and subjecting the system to strong
perpendicular magnetic fields $B$, results in quantized Hall resistances when magneto-transport measurements are performed. This phenomena is called the quantized Hall effect (QHE)~\cite{vKlitzing80:494}. The quantized Hall
resistivity and longitudinal resistivity are often explained as a
result of Landau quantization at high magnetic fields. Quantized
levels are called Landau levels (LLs) which are highly degenerate,
and the energy of the LLs are given by $E_n=(n+1/2)\hbar\omega_c$,
where $n$ is the level index, $\hbar$ $(=h/2\pi)$ is Planck's
constant and $\omega_c={eB/m^*}$ represents the cyclotron frequency
of an electron with an effective mass $m^*$ ($=0.067m_e$ for
GaAs/AlGaAs heterostructures).

After the discovery of QHE, considerable attention
is diverted to the unexpected transport properties of 2DESs. The current transport in a 2DES is usually described by two main pictures, namely the bulk~\cite{Kramer03:172} and the edge
pictures~\cite{Chklovskii92:4026, Halperin82:2185, Buettiker86:1761,
siddiki04:195335}. In the bulk picture one assumes an infinite homogeneous 2DES, where the current flows from the
bulk. Transport properties are essentially determined by impurity
scattering and localization arguments are dominant~\cite{Kramer03:172}. Conversely, in the edge picture a finite system is considered and the
current is assumed to be confined near the edges of the sample. The transport is
mainly due to back-scattering suppressed, one dimensional ballistic channels. The properties of these channels are widely
investigated in many context~\cite{Beenakker89:2020,Levkivskyi08:045322,McClure09}.
However, if one takes into account the direct Coulomb interactions the 1D channel picture is modified. Due to the presence of a
perpendicular magnetic field, the 2DES has a peculiar density of states (DOS) distribution as a function of energy (and position). The peculiarity of the DOS result in unusual screening
properties~\cite{Chklovskii92:4026,Wulf88:4218,Siddiki03:125315}.
Such that, if the Fermi level falls between two adjacent LLs,
electrons cannot be redistributed. Therefore, they do not contribute
to screening locally. The region exhibiting poor screening
properties is called incompressible and it behaves \emph{like} an
insulator. For further references, it is useful to define a
dimensionless parameter: The filling factor $\nu=n_{\rm
el}/n_{\phi}$, where $n_{\rm el}$ and $n_{\phi}$ are the number densities of
the electron and the magnetic flux for a given area $A$, respectively. The
local filling factor $\nu(x,y)=n_{\rm el}(x,y)/n_{\phi}$, measures
the number of fully occupied Landau levels (locally), where $x$ and
$y$ denote the coordinates in real space. Here it is assumed that
the $B$ field is homogeneous both in space and time. Therefore, if
the region is incompressible, the filling factor equals to an
integer. Whereas, if the Fermi level is pinned to one of the Landau
levels, \emph{i.e.} the level is partially occupied ($\nu=$
non-integer), electrons can be redistributed. Hence, the region is
compressible and it behaves \emph{like} a metal. Starting from late
80's, the formation of these peculiar strips have been investigated
theoretically, considering different approaches and approximations.
One of the most appreciated scheme is the so-called Chklovskii
picture~\cite{Chklovskii92:4026}, where a Thomas-Fermi mean field
approximation is considered to obtain electrostatic properties of
the 2DES at the edges of the system in a non-self consistent manner
at zero temperature~\cite{Chklovskii92:4026,Chklovskii93:12605}.
Subsequently, K. Lier and R. R. Gerhardts have investigated the
properties of these strips at finite temperatures, performing
self-consistent (SC) calculations~\cite{Lier94:7757}. In this work,
the deviations from the Chklovskii picture are already shown both in
spatial distribution and in calculation of the widths of the strips.
Afterwards, it is shown that the incompressible strips are much
narrower than the predictions of the pioneering
work~\cite{Oh97:13519}, considering the effects resulting from the
growth directions. The reports mentioned above are all based on the
Thomas-Fermi approximation, which explicitly assumes that the
overall potential profile varies slowly on the quantum mechanical
length scales, \emph{e.g.} on the scale of the wave length. Hence, the
finite width of the wave functions is neglected and quantum
mechanical properties of the electrons are ignored such as finite
tunneling rates \emph{etc}. However, if an incompressible strip is
formed, a strong potential variation exists, therefore, if the strip
width becomes comparable with the wave extent, TFA becomes
questionable~\cite{siddiki04:195335}, which we also address in the
present work. The first order quantum mechanical effects were
included to the self-consistent calculations within a mean-field
Hartree approximation, where the results show that the
incompressible strips vanish in certain magnetic field
intervals~\cite{siddiki04:195335,Suzuki93:2986}. These results have
important consequences on the transport taking place at the 2DES
both theoretical~\cite{SiddikiEPL:09} and experimental
wise~\cite{Ahlswede02:165,Afifcurentexp}.

In this paper, we investigate the non-self-consistent Thomas-Fermi
approach of Chklovskii, Shklovskii and Glazmann (CSG) in detail to clarify the effects stemming from experimental conditions. For this purpose, the electron
density distribution and the widths of incompressible strips are
calculated self-consistently considering different sample preparation conditions, namely by gates, by chemical
etching processes and a superposition of these procedures. We first summarize the analytical approach of the
mentioned work and clearly state the assumptions made. In the next
step, we present the results of a fully self-consistent calculation
performed on a 3D crystal and compare the electron densities with
the analytical results. Utilizing the CSG formalism the widths of the incompressible strips are
calculated also considering spin degree of freedom, disorder and quantum
mechanical effects. In this work, we explicitly show that more than one incompressible strip with different filling factors cannot form at typical Hall bar geometries due to the
steepness of the total potential at the edges. In a final section, we discuss the
different regimes of the edge profiles considering gated, etched and
trench gated samples.
\section{The Edge-states}
Since the quantized Hall effect is related to the existence of the
``edge-states", it is useful to clarify the definition of these
states at different pictures, while the notation
becomes confusing for a non-expert. One first encounters the
notation edge-states in an early work by
Halperin~\cite{Halperin82:2185}, where he discusses finite size
effects arising from the real sample structure, \emph{i.e.} the
confinement potential. Such an approach is fairly different than the
conventional bulk picture, which neglects the boundary effects all
together~\cite{Kramer03:172,Laughlin81} and attributes the quantized
Hall effect to gauge invariance and localization of the electronic
states. Even a brief discussion of the bulk theory, hence localization picture, is far beyond the scope the present work. Therefore, we avoid such a discussion. However, the basic idea
in a superficial manner is to localize electrons to bulk states due
to disorder. Hence, these electrons cannot contribute to transport
and therefore the Hall conductance is quantized whereas, the longitudinal
conductance vanishes.
\begin{figure}[h!]
\begin{center}\leavevmode
\includegraphics[width=0.7\textwidth]{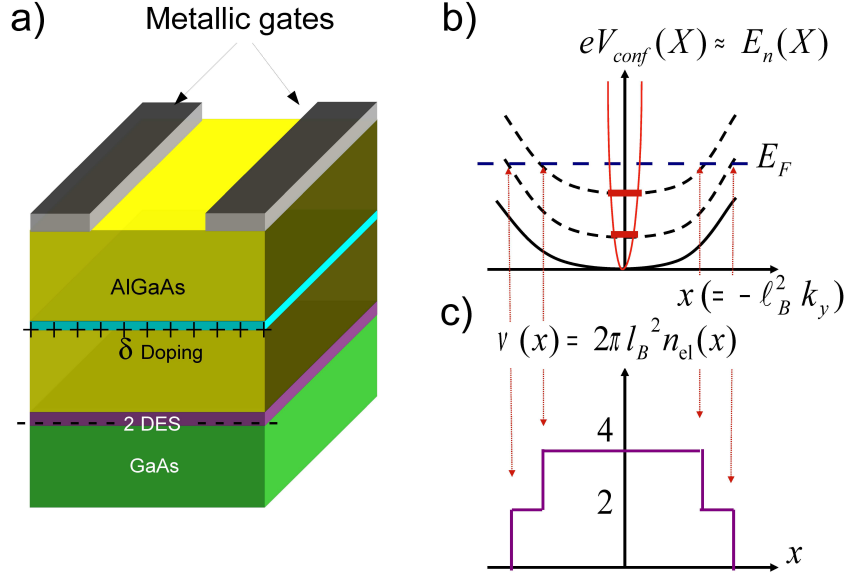}
\caption{(a) Schematic presentation of the crystal. (b) The curves depict the confinement energy (thick black curve), the Landau levels
(broken lines following the confinement potential) and the Fermi energy
(horizontal dashed line), respectively. Red thin parabolic curve and red thick lines
depict the magnetic confinement and the position dependent Landau levels, respectively.
(c) The estimated plot of filling factor as a function of position from (b).
}\label{potential}
\end{center}
\end{figure}
\subsection{The Halperin and Landauer-B\"uttiker type edge-states}
In contrast with the bulk picture, one neglects the disorder
effects and considers the finite size of the system instead, it is
observed that the single particle Landau levels are bent at the
edges due to the confinement potential. Fig.~\ref{potential}(a) presents a typical crystal structure and
Fig.~\ref{potential}(b) shows the confinement due to the remote (homogeneous) donors (thick black curve) and the magnetic
confinement of a single particle (broken lines) together
with the position dependent Landau levels (red lines following
the confinement potential). Such a plot corresponds to the single
particle Hamiltonian given as,
\begin{equation}
H_i=\frac{1}{2m^*}(\vec{p}_i(\vec{r}_i)+\frac{e}{c}\vec{A}(\vec{r}_i))^2+V_{\rm
conf}(\vec{r}_i),
\end{equation}
where we assume a translational invariance in $y$ direction and
consider the Landau gauge to define the $B$ field in the $z$
direction, \emph{i.e.} $B=(0,0,B)=\nabla\times A(y,-x,0)$. Note that
the horizontal axes in Fig.~\ref{potential}(b)-(c) are both in real
$x$ and momentum $k_y$ spaces, where the center coordinate $X_0$ is
related to $k_y$ by $X_0=-l_B^2k_y$, and $l_B=\sqrt{\hbar/eB}$ is
the magnetic length. Here, $k_y$ is the quasi-continuous momentum in
$y$ direction. The crucial assumption to have such an energy
dispersion is to have a smooth confinement potential, hence, one can
replace the $X_0$ dependent energy eigenvalues by a constant given
by the local value of the confinement potential (energy),
\emph{i.e.} $E_n(X_0)=E_n+V(X_0)$. This enables us to fix the
spatial (and energetic) distribution(s) of the edge-states for a
given Fermi energy $E_F$, (dark) dashed horizontal broken line. When
it comes to transport, it is easy to conclude that the current is
carried by the edge-channels while there are available states at the
Fermi energy once $E_F$ cuts through the Landau levels. At
equilibrium, the number of edge channels carrying positive momentum
$k_y$ (namely forward movers) equals to number of channels carrying
negative momentum $-k_y$ (namely backward movers) and as a result of
every channel can only carry a unit of conductance (the velocity is
canceled due to the density of states in 1D), net current is zero.
To impose a finite current, one has to increase the chemical
potential energy of forward movers by an amount of $\mu_{SD}$, hence
the potential (energy) difference between the probe (side) contacts
is now quantized to the number of edge channels. A detailed
description can be found in Ref.~\cite{Davies}. [Such a transport
scheme is highly out of equilibrium, since one injects electrons
which have considerably higher energies than the electron sea, and
the unoccupied states carry the non-equilibrium \emph{excess}
current.] Moreover, the injected electrons do not induce a Hall
potential which is spread across the sample; the spatial
distribution of the electrochemical potential is overlooked and the
results deviate strongly from the experimental
findings~\cite{Ahlswede01:562,Yacoby04:328}. The above formalism is
known as the Landauer-B\"uttiker (LB)
picture~\cite{Buettiker86:1761}, and these edge-states are named as
(1D and ballistic) Landauer-B\"uttiker edge states (LBES). Since we
neglected disorder and the bulk of the sample is completely
incompressible, as a result, no back-scattering can take place,
therefore the longitudinal resistance vanishes which is accompanied
by the quantized Hall resistance. However, such a picture is not
adequate to explain the transitions between the quantized Hall
plateaus observed experimentally. Hence, one should allow
\emph{some} scattering due to the impurities and introduce bulk
states, where electrons are orbiting around hills and valleys of the
disorder potential. Such a description is so far so good only if one
can bare the fact that the resulting electron density distribution
is stepwise and highly unstable (see Fig.~\ref{potential}(c)), due
to the strong Coulomb force between electrons. In fact,
chronologically, Halperin introduced the concept of ``edge-states''
even earlier than the mentioned work above~\cite{Halperin82:2185}.
However, there the edge-states are obtained for a system which
presumes infinite walls at the physical edges of the system. For
such a system, the solution of the Schr\"odinger equation can be
obtained analytically. The wave functions are described by parabolic
cylindrical functions, whereas the Landau levels depending on the
center coordinate $X_0$ are \emph{only} bent at the edges, in
contrast to LB picture where potential varies smoothly in the scale
of $l_B$. An important implication of such infinite walls is that
the energy dispersion $E_n(X_0)$ in no longer equidistant, $\Delta
E(X_0)=E_{n+1}(X_0)-E_{n}(X_0)\neq\hbar\omega_c$. Utilizing
Bohr-Sommerfeld quantization one can obtain the dispersion at the
edges~\cite{Gerhardts:unpub} or by considering WKB approximation one
has~\cite{Avishai:09}
\begin{equation}
E_n(X_0)/\hbar\omega_c\approx (4n+3)+\frac{4X_0}{\sqrt{ \pi}}
\prod_{p = 0}^{n} (1+ \frac{1}{2p})
\end{equation}
Hence, equidistant quantization at the bulk, \emph{i.e} $X_0$ far
from the infinite walls, is recovered.

It is important to emphasize that, the approach of Halperin is rather different than the B\"uttiker one,
 in the first case infinite walls are assumed at the edges and the eigenfunctions $\psi_{n,X_0}(x,y)$ satisfy the
 condition $\psi_{n,X_0}(x\rightarrow \rm{boundary},y)=0$ and the external potential is zero in the bulk. In the
 latter case, the external potential varies smoothly and the eigenfunctions satisfy $\psi_{n,X_0}(x\rightarrow \inf,y)$.
 Clearly, these two approaches assume different boundary conditions and impose different criteria on the external potential.
\subsection{The Chklovskii edge-states}
The discrepancy due to the stepwise behavior of the density profile
is cured, if one includes the classical electron-electron
interactions (direct Coulomb), which was discussed even earlier than
Chklovskii \emph{et al} by R. R. Gerhardts and his
co-workers~\cite{Wulf88:4218,Wulf88:162} and by A. M.
Chang~\cite{Chang90:871}. However, we focus our discussion on the
elegant analytical investigation of Chklovskii \emph{et al}. They
studied the electrostatics of the edge channels and provided an
analytic expression for the widths of
the incompressible strips depending on the filling factor. In their work, 2DES is
considered to be formed in a $GaAs/Al_xGa_{1-x}As$ heterostructure
(in fact in their work the 2DES is formed at the interface of a
semiconductor and air), where the fluctuations at the donor
concentration is neglected and the donor density set equal to
electron density far from the boundaries. They create the boundary
of 2DES by applying a negative voltage $-V_g$ to the metallic gate
on the surface assuming a half-plane
geometry. 
The solution
assumes a translational invariance in the $y-$direction.
The gate potential determines the width of depleted strip ($l_d$)
and is given by
\begin{equation}\label{ldep}
l_d=\frac{V_g \epsilon}{4\pi^2 n_0 e} ,
\end{equation} where $n_0$
is the surface number density of the homogeneous donor layer. In
this model a capacitor is considered where metallic plates are in
the same ($z=0$) plane together with the donor \emph{and} electron
layers. The metal plates are assumed to be separated from each other
by an uniformly charged insulator. One of the metal plates is
exposed to a negative gate potential and the other metal plate is
considered as grounded. Thus, the gate potential pushes the
electrons to the grounded plate where the 2DES is formed. Since, the
$z<0$ half-space is occupied by a semiconductor with a high
dielectric constant ($\epsilon>>1$) the Laplace equation can be
solved. As a result, they have obtained the following equation for
the spatial distribution of the electron density in the metallic
\emph{like} strip begins at the end of the depletion region:
\begin{equation}\label{eq:nelx}
n(x)=n_0\sqrt{\frac{x-l_d}{x+l_d}} ,  \,  |x|>l_d,\\
\end{equation}
which reaches to its bulk value $n_0$ far away from the boundary. Note
that the electron density given by the above equation is eligible in
the absence of magnetic field, only considering in-plane gates which
reside at the plane of 2DES.

Next, they consider single particle interactions in 2DES and the
ladder like behavior is replaced by smoother density distribution
that present narrow constant density regions, i.e incompressible strips. They propose that co-existing (and adjacent)
compressible and incompressible regions are formed in the presence
of strong magnetic fields. They assume that the compressible regions
behave like a metal so screening is perfect and electrostatic
potential is \emph{completely} constant, whereas in the
incompressible regions, screening is very poor and the system
\emph{behaves} like an insulator.
As a result of analytic calculations, they derived the generalized
expression for the strip widths assuming any integer number $k=1, 2,
. . ., M$: 
\begin{equation}\label{aklnlk}
a_k^2=\frac{2\epsilon\Delta{E}}{\pi^2 e^2 dn/dx|_{x=x_k}},
\end{equation}
where $dn/dx|_{x=x_k}$ is the derivative of electron density with
respect to spatial coordinate $x$ calculated at the center of the
strip, \emph{i.e} at $x_k$, and the single particle energy gap $\Delta E$ is
denoted by $\Delta{E}$. It is very important to note that, a
perfect metal is assumed for the compressible regions ($\nu=$
non-integer, $r_s=0$) and a perfect insulator is assumed for the
incompressible strips ($\nu=$ integer, $r_s=\infty$), such an
assumption cannot be justified for a 2DES which is formed at the
interface of two semiconductor materials for sure. A simple
calculation of the screened potential
\begin{equation} V_{\rm
scr}(q)=V_{\rm ext}(q)/\epsilon(q)
\end{equation}
for a given external potential $V_{\rm ext}(q)$ via dielectric
function \begin{equation} \epsilon(q)=1+\frac{2\pi
e^2D_0}{q\epsilon}
\end{equation}
 together with the constant
density of states ($D_0$) of a 2D system already shows that, if the
$D_0$ is not infinite (like in a metal), such an approximation
fails. These compressible and incompressible strips are known as the
Chklovskii edge-states, which are no longer 1D channels. However, as
we discuss later once more, the vanishing longitudinal resistance is
attributed to the absence of backscattering and one obtains the
quantized Hall plateaus simply by counting the number of
\emph{compressible} strips. Hence, 1D LB edge states are replaced by
current carrying compressible strips, separated by insulating
incompressible strips that suppress back-scattering.

In the original work, spin degree of freedom is neglected, hence,
the energy gap equals to $\hbar \omega_c$. Therefore, all integer
(odd or even) filling factors assume the same energy gap. However,
in our calculations, we consider spin splitting in a rather
phenomenological manner and take into consideration Zeeman splitting
by assuming an enhanced effective Lande-$g^*$ factor. Actual value
of $g^*$-factor depends on whether exchange component of Coulomb
interaction is taken into account or not. Since the exchange interaction
depends on the spin polarization of the system, $g^*$-factor varies depending on the magnetic field.
Enhancement of effective $g^*$ factor is already investigated by Manolescu and
Gerhardts~\cite{Manolescu} almost two decades ago and it has been
shown that local spin polarization leads to wider spin polarized
incompressible strips due to spatially enhanced Zeeman gap. However,
here for simplicity we assume that $g^*$ factor does not depend on
$B$. Since at the magnetic field intervals we are interested in, the incompressible stripes are
already narrow and spin polarization becomes small, the effective Zeeman splitting will become
less significant therefore $B$ dependency of $g^*$ will bring only small quantitative corrections.
For a detailed study of spin effects on incompressible strips we refer to a recent spin density
functional theory calculation considering bulk $g^*=-0.44$~\cite{Ihnatsenka06,Ihnatsenka07} and
predicting collapse of the Zeeman gap due to small $g^*$. Moreover, under typical experimental
conditions (\emph{e.g.} $T$=1 K, $B$=5 T and low-mobility samples), the gap already closes and
the odd incompressible strip
disappears. However, due to exchange-correlation effects it is explicitly shown that the incompressible
strips are indirectly enhanced~\cite{Ihnatsenka06,Ihnatsenka07}. In principle one can fit the $g^*$
factor using the widths of odd incompressible strips from the experiments~\cite{Ahlswede02:165,Ahlswede01:562}.
This experiments showed that exchange and correlation effects enlarge the energy gap of the LLs which
corresponds to odd filling factor as if the $g^*$ factor value
is similar to $5.2$. However, since our aim is to show qualitatively that the odd (also even) integer
incompressible strips collapse due to the strength of the confinement potential, we use a constant $g^*$,
which nevertheless describes the system properly.
 Therefore, if one takes into account the Zeeman splitting, the
energy gaps are given as
\begin{equation}\label{ak1_2}
\triangle{E}=\left\{ \begin{array}{ll}
g^*\mu_BB&\quad , \, \nu=odd \\
\hbar \omega_c-g^*\mu_BB & \quad, \, \nu=even\\
\end{array} \right.
\end{equation}

Another handicap of the Chklovskii model is due to the considered
geometry, namely the half-plane, which is cured in the subsequent
work~\cite{Chklovskii93:12605}. However, the 2DES, gates and donors
reside on the same plane; but this is not reasonable for a gated and
shallow etched defined realistic samples. As an exception, deep
etched samples can still be approximated by the Chklovskii model in
calculating the strip positions, if the gate potential is set to
-0.75 V, corresponding to mid gap energy of GaAs. However, the
widths are usually over
estimated~\cite{siddiki04:195335,Ahlswede02:165}. In our
calculations, we consider a sample which has the dimensions of a
$2.4$ $\mu$m $\times$ $2.4$ $\mu$m unit cell, where side contacts
also have realistic dimensions ($\sim 300$ nm in width and $\sim
2400$ nm in length, residing on both sides). 3D Poisson equation is
solved self-consistently for this geometry (see
Fig.~\ref{potential}(a)), using the successful 4$^{th}$ order grid
technique.
In three dimensions, evaluating both the potential and charge
densities comparatively is the essence of 4th order grid
technique~\cite{Andreas03:potential}. At this technique, initial
potential (gate potential) and charge (donor) density are given for
a grid point, and using Poisson equation, the nonuniform potential
on the outer boundary (due to the geometry) and charges are
iteratively calculated on the nearest neighbors grid points. This
technique is implemented to many other similar
structures~\cite{Andreas03:potential,Sefa08:prb,Engin09:physica}.
Thus, we numerically obtain the electron density and total
electrostatic potential distributions for a given crystal structure
and lithographic pattern. By applying different gate voltages to the
metallic gates and changing the etching depths of the structure we
control the electron density distribution and also manipulate the
depletion length.
\begin{figure}[h!]
\begin{center}\leavevmode
\includegraphics[scale=0.27]{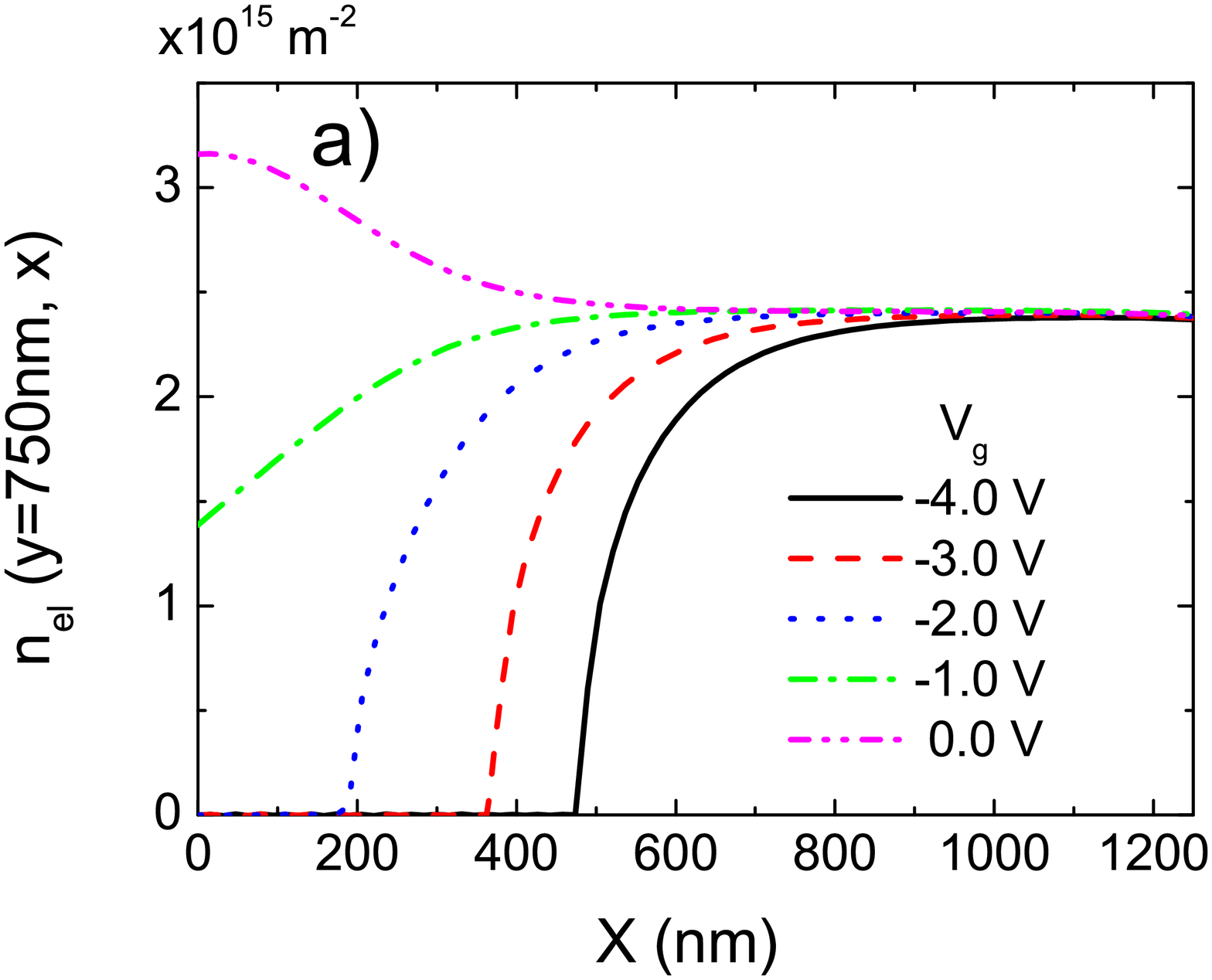}
\includegraphics[scale=0.27]{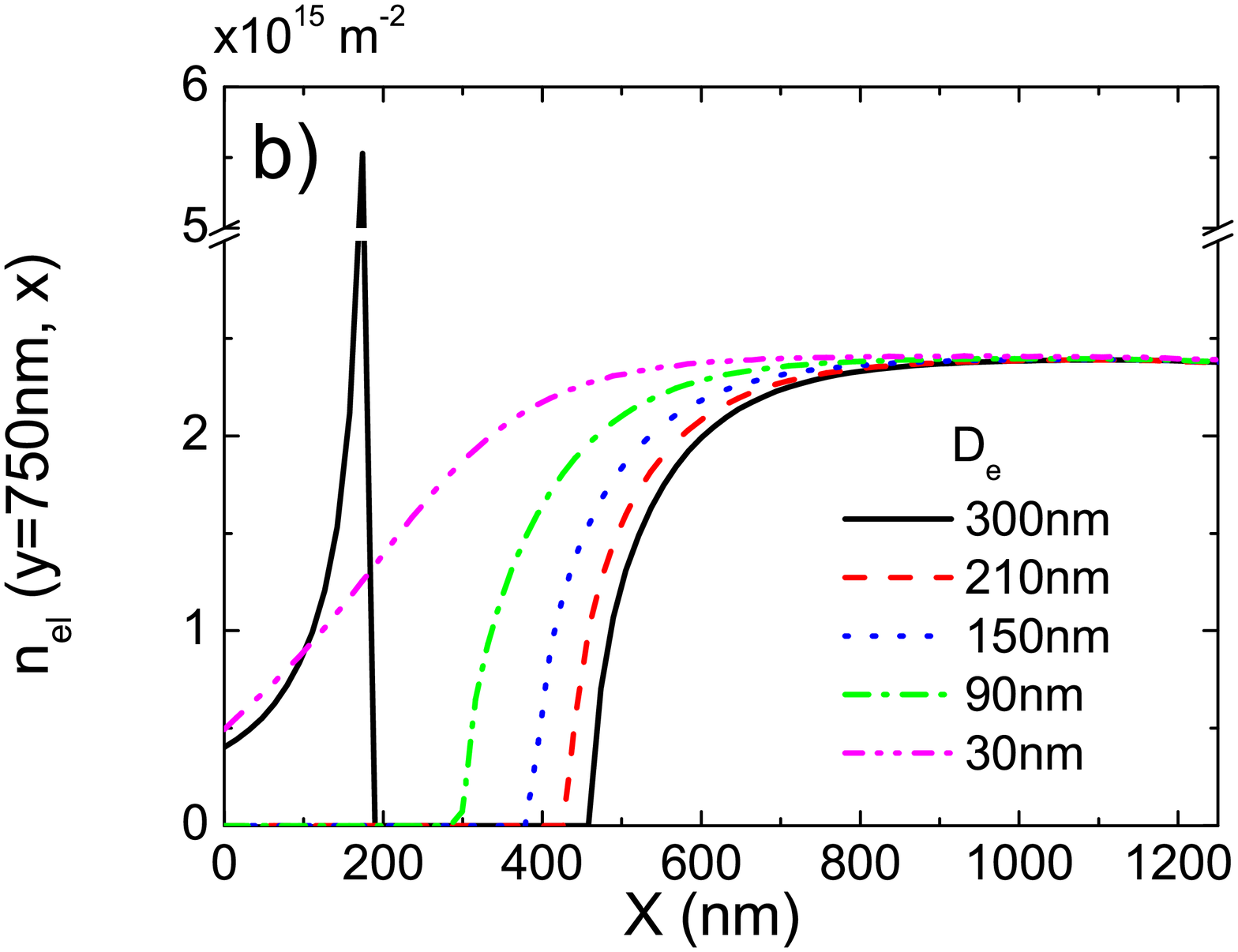}
\includegraphics[scale=0.27]{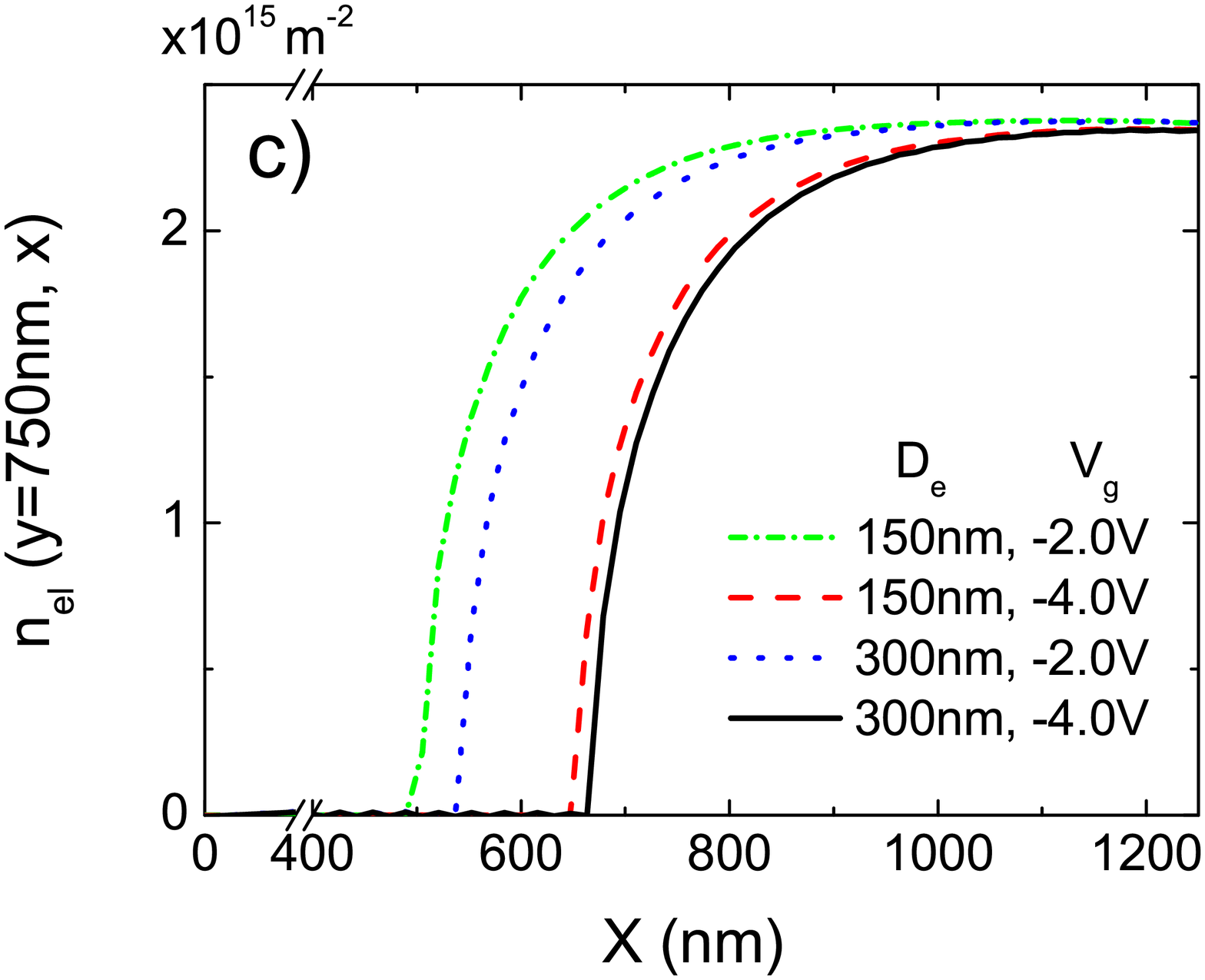}
\caption{(Color online) Electron density distribution of a realistic
sample as a function of position up to the bulk, starting from the
physical boundary. (a) The gate defined sample, (b) the etch defined
sample and (c) the trench-gated sample. The heterostructure
parameters, such as the depth of the electron layer, are fixed
whereas the gate potential and etching depth are
varied.}\label{density}
\end{center}
\end{figure}
\subsection{Full electrostatics of the crystal}
In this section we show the results of our self-consistent
calculations, considering gate, etched and trench gated structures
while varying the depth of the 2DES and etching together with the
gate potentials. The main outcomes of our simulations are presented
in Fig.~\ref{density}, where we show the electron density
distribution as a function of lateral coordinate $x$ considering a
gate defined structure (a), an etched sample (b) and a trench gated
sample (c). The extent of depleted regions is manipulated by
applying different gate potentials and/or etching depths.
Fig.~\ref{density}(a) presents density variation for typical gate
potentials, note that the gates are at the surface and the surface
potential is pinned to the mid gap of GaAs, \emph{i.e.} $v_S=-0.75$
V. Therefore, zero gate potential (double doted-dashed curve)
behaves as a positive potential, relative to the surface, hence more
electrons are accumulated beneath the gate. This is the first
difference between our 3D calculations and the in-plane gated
Chklovskii geometry, there even a small gate voltage generates a
depleted strip. In contrast, we can obtain the pinch-off voltages of
the gates in a self-consistent manner. The first negative potential
bias is $-1.0$ V (note that this value is effectively -.25 V), where
the density of the 2DES starts to be reduced below the gates. Only
if a larger negative potential is applied to the gates, one observes
an electron depleted region (dotted line) $V_g=-2.0$ V, with a
depletion length of $190$ nm. One can clearly see from the gate
potential dependence that the $l_d$ does not scale linearly with the
applied gate voltage as assumed in equation~(\ref{ldep}). In our
calculations, we consider a heterostructure geometry as shown in
Fig.~\ref{potential}(a). The total hight of the sample is 440 nm,
where the 2DES lies 288 nm below the surface and the delta doped
donors are located 165.5 nm below.
Next, we consider the etching defined samples. We investigate the
electron density distributions of the sample for varying etching
depths ($D_e$) and the results are showed in Fig.~\ref{density}(b).
If a tiny layer is removed from the surface (\emph{i.e.} $D_e=30$
nm, double dotted-dashed pink line) no electron depleted region
occurs, likewise the low gate bias. Once the etching depth becomes
larger than $90$ nm, an electron depleted strip forms. However, note
that the width of the depleted region is much larger when compared
to the gated sample. One can observe that the etched sample in
Fig.~\ref{density}(b) the electron density changes with etching
depths, as expected. Most interestingly, if one etches the crystal
below the 2DES layer ($D_e>288$ nm, black solid line), side surface
charges become visible. This is the regime where analytical
calculations of Gefland and Halperin~\cite{Halperin94:etchedge} fit
better to the numerical results. These side charges push the 2DES
even stronger than an in-plane gate, since they are spread all over
the side surface. Another way to define narrow constrictions is the
so-called trench gating. Here, first some layer of crystal is
removed starting from the surface and then the metallic gate is
deposited to this etched region. Of course, this procedure is much
more complicated than either by gating or etching, however, has the
advantage to generate steeper confinement potentials due to etching,
and also one can control this steepness by applying a potential on
the trench gates. Such structures are preferred usually in
interference devices~\cite{Goldman05:155313,goldman07:e/3}, since a
full control of the transmission parameters are
necessary~\cite{Engin09:physica}. Fig.~\ref{density}(c) depicts the
resulting electron density considering trench gated samples, which
leads wider depleted regions compared to other boundary defining
techniques. Applying negative gate voltages to the etched sample,
pushes the (side) surface charges to the bulk of the sample and the
peak in the Fig.~\ref{density}(b) vanishes, since now the side
charges are captured by the metallic gates. It is apparent that the
derivative of the density profile (\emph{i.e.} the slope) is much
larger than the gated structures, hence, according to the
equation~(\ref{aklnlk}) the resulting incompressible strip will be
much narrower compared solely defined by gates.

\begin{figure}[h!]
\begin{center}\leavevmode
\includegraphics[scale=0.3]{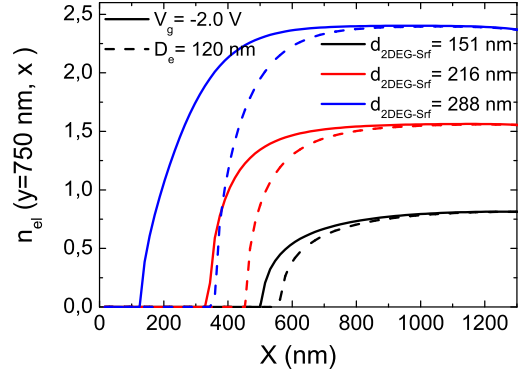}
\caption{(Color online) Spatial distribution of the electron
density with different heterostructure parameters. Solid line
corresponds to gated samples and dashed line etched samples. A negative voltage -2.0 V is applied to the gates, whereas
for the etched samples, the depth of etching is fixed to 150 nm. The distances between electron gas and surface
$d_{2DEG-Srf}$ are taken to be 151 nm, 216 nm, 288 nm. The corresponding depths of the donor layers are $86.4$ nm, $122.4$ nm and
nm, $165.6$ nm, respectively. Similar calculations are also performed considering $d_{2DEG-Srf}=$ 180nm and 266 nm, where
$d_{donor-Srf}=$ 100.8 nm and 151.2 nm, however, the tendencies are left unchanged.}\label{different_d}
\end{center}
\end{figure}

In a next step, we also investigated various electron layer depths from the
surface ($d_{2DEG-Srf}$), while also varying the donor layer depth from the
surface ($d_{donor-Srf}$), as shown in Fig.~\ref{different_d}, while fixing the gate potential or etching depth.
The first observation is that the distance between the surface and the electron layer effects the bulk
electron density considerably. Once the electron layer is further away from the surface, i.e. the gates,
the electron density increases almost exponentially, since the electrostatic potential decreases rapidly with the distance.
As an important note, we point that the gate defined edges provide a smoother potential profile near the physical boundaries
of the sample when compared to etch defined samples. By performing calculations considering various depths of the 2DES and
the donor layer we provide the following empirical formula to estimate the depletion length:

\begin{equation}
l_d=\frac{\epsilon}{e n_0} (\frac{d_{2DEG-Srf}} {d_{sample}})(c_1
V_g)[1-\frac{d_{donor-Srf}}{10 a_B^*} exp({-\frac{\epsilon
a_B^*}{c_{2} e} \frac{d_{donor-Srf}}{d_{2DEG-Srf}} V_g})],
\label{eq:ldampricV}
\end{equation}
where $c_1$ is a normalization constant that that depends on the applied
gate potential given by  $c_1 V_g\approx0.24V$ condition, whereas
$c_2\approx84$ is a dimensionless constant as a free fit parameter.

Another important conclusion that can be made from our self-consistent calculations is that larger gate potentials essentially yield steeper confinement at the edges. Performing a suitable fitting
results in the following form of the electron density
\begin{equation}
n_{\rm el}(x)=(1-e^{-(x-l_d)/t})n_0,\label{eq:nexp}
\end{equation}
where $l_d$ determines the electron poor region width just in front
of the gate and $t$ is a parameter related with the slope of
electron density. We observed that $t$ is the order of $10 a_B^*$
for smooth electron density, consistent with the non-self-consistent
calculations.

We also performed a similar fitting process to the etched defined samples, where the non-linear relation between etching
depth and depletion length is observed once more. Interpolating the
relation between the etching depth and depletion length over various
sample parameters yield the form
\begin{equation}
l_d=\frac{\pi}{n_0 {a_B^*}}
(\frac{d_{2DEG-Srf}}{d_{sample}})[c_3-{\frac{d_{donor-Srf}}{c_3
a_B^*}} exp({-\frac{D_e}{10 a_B^*}})],  \label{eq:ldampricD}
\end{equation}
where the constant is calculated as $c_3\approx4.5$. A similar
density profile given in equation~(\ref{eq:nexp}) fits perfectly to
our self-consistent results.
\begin{figure}[h!]
\begin{center}\leavevmode
\includegraphics[scale=0.3]{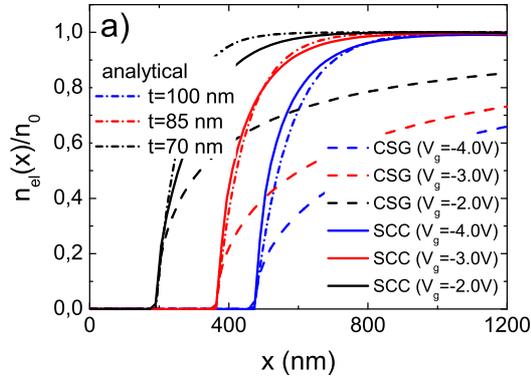}
\caption{Comparison of densities which are obtained by analytical
formulation, self-consistent calculation (SCC) and Chklovskii
formalism (CSG).}\label{analytical}
\end{center}
\end{figure}
This is shown in Fig.~\ref{analytical}, where the density profiles
are calculated using the self-consistent scheme (solid lines), the
empirical formula - Eq.~\ref{eq:nexp} (dash-dotted) and the
analytical formula provided by Chklovskii \emph{et al}, i.e.
Eq.~\ref{eq:nelx} (dashed line). One observes fairly good overlap
between the self-consistent calculations and Eq.~\ref{eq:nexp},
whereas the Chklovskii prediction fails by a gross amount, both
qualitatively and quantitatively.

So far we have shown by solving the full electrostatics that, the analytical non-self-consistent
Thomas-Fermi approach predicts rather different density profiles
compared to our results. Now the question is how important are these
deviations? In fact, when calculating the strip widths, these
differences become much more emphasized, since the widths ($a_k$)
strongly depend on the derivative of the electron density. Next we
present our numerical results calculated within the Chklovskii
picture, which show that co-existence of many incompressible strips
with different integer filling factors are barely possible.

\subsection{Strip widths}

\begin{figure}[h!]
\begin{center}\leavevmode
\includegraphics[scale=0.3]{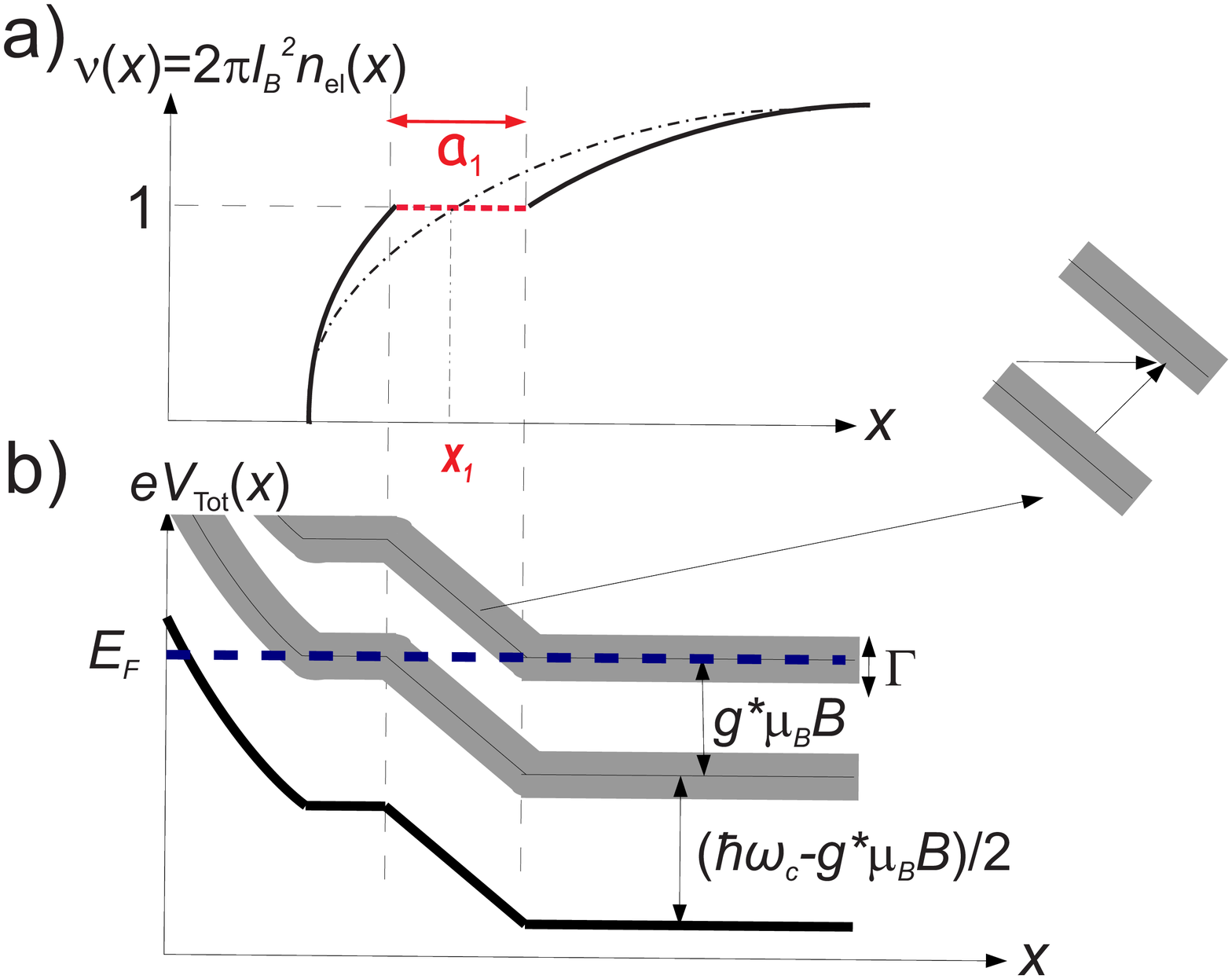}
\caption{(a) Schematic representation of the formation of
incompressible strip $\nu=1$ at $x_1$. (b) The variation of the
total potential with respect to spatial coordinates. The solid thick
line and shifted thin lines represent background potential and
Landau levels, respectively. Grey region indicates the enlargement
of Landau Levels due to disorder, by an amount of $\Gamma$, whereas
the dashed horizontal line is Fermi energy. }\label{schematic_IS}
\end{center}
\end{figure}

\begin{figure}[h!]
\begin{center}\leavevmode
\includegraphics[scale=0.3]{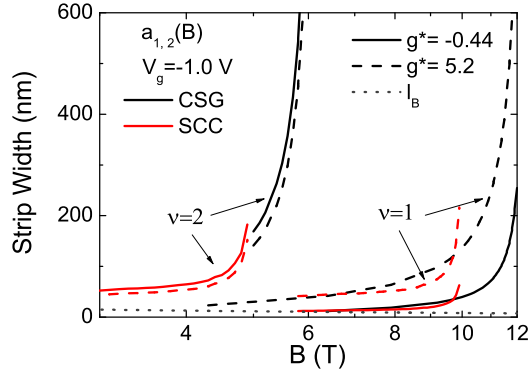}
\caption{Incompressible strip widths of $\nu=1$ and $\nu=2$
considering different Lande-$g^*$ factors. The strip widths with
black lines correspond to the Chklovskii formalism and red lines
correspond to the self-consistent calculation.}\label{IS_widths}
\end{center}
\end{figure}

Once the electron density distribution is obtained from
electrostatics, it is an easy task to locate the center of the strip
for a given $B$ by the relation $\nu(x_k)=2\pi l_B^2n_{\rm el}(x_k)$
and the width of the strip from Eq.~(\ref{aklnlk}), as shown in
Fig.~\ref{schematic_IS}(a) for $k=1$. Here the thin curve depicts
the electron density distribution when magnetic field is absent.
Fig.~\ref{schematic_IS}(b) represents the potential energy
distribution as a function of position. As depicted in
Fig.~\ref{schematic_IS}(b) one observes a potential drop when
incompressible strip is formed.
Similar to Chklovskii's analysis, however, utilizing the
self-consistently obtained electron density distribution
(Eq.~\ref{eq:nexp}), we obtain the following expression to determine
the widths of the incompressible strips
\begin{equation}\label{ourwidth}
a_{1,2}^2= \frac{4 a_B^* \alpha_{1,2}}{\pi}
\frac{t}{e^{-(x_1-l_d)/t}},
\end{equation}
where  $a_1$ and $a_2$ corresponds to incompressible strip widths of
$\nu=1$ and $\nu=2$, respectively and $\alpha_1=(g^*\mu_BB)/\hbar
\omega_c$, $\alpha_2=(\hbar \omega_c-g^*\mu_BB)/\hbar \omega_c$ are
the dimensionless energy parameters. In Fig.~\ref{IS_widths}, we
show the widths of the incompressible strips as a function of
magnetic field calculated within the Chklovskii formalism and our
self-consistent calculations.

For the moment we neglect the finite temperature effects and
calculate $a_1$ and $a_2$ depending on the field strength while
considering a gate defined sample under the -1.0 V gate bias with
$n_0=3\times 10^{15}$ m$^{-2}$ donor concentration. Note that, at
this voltage, the 2DES is still not depleted and the electron
density varies smoothly. In Fig.~\ref{IS_widths}, we see the
incompressible strips widths obtained within the CSG
picture~\cite{Chklovskii92:4026} (black curves) and our SCC
calculations (red curves). We calculate the incompressible strips
with two different $g^*$ factors. As mentioned before, we consider
either the bulk $g^*$ factor or an exchange enhanced one (taken to
be 5.2, similar to the experimental values~\cite{Khrapai:05}).
Although in our sample, the incompressible strips widths which are
calculated by SCC are finite with increasing magnetic field, the
strip widths increase unrealistically within the CSG calculation,
since a half-plane geometry is assumed in the latter work. The real
samples as our sample assume boundaries which take the electron
density in a finite region. We see that, the Chklovskii picture
yields artificially wide strips due to the incorrect modeling of the
density profiles both for $|g^*|=0.44$ and 5.2. Ihnatsenka and
Zozoulenko are compared the results of Chklovskii \emph{et al} and
their density functional theory calculations considering the
formation of the compressible strips~\cite{IhnatsenkaZozoulenko06},
where it is shown that strip widths are narrower than CSG picture.
Apparently, our results are considerably more consistent with
Ref.~\cite{IhnatsenkaZozoulenko06} (cf. Fig.~\ref{IS_widths}).

\begin{figure}[h!]
\begin{center}\leavevmode
\includegraphics[scale=0.5]{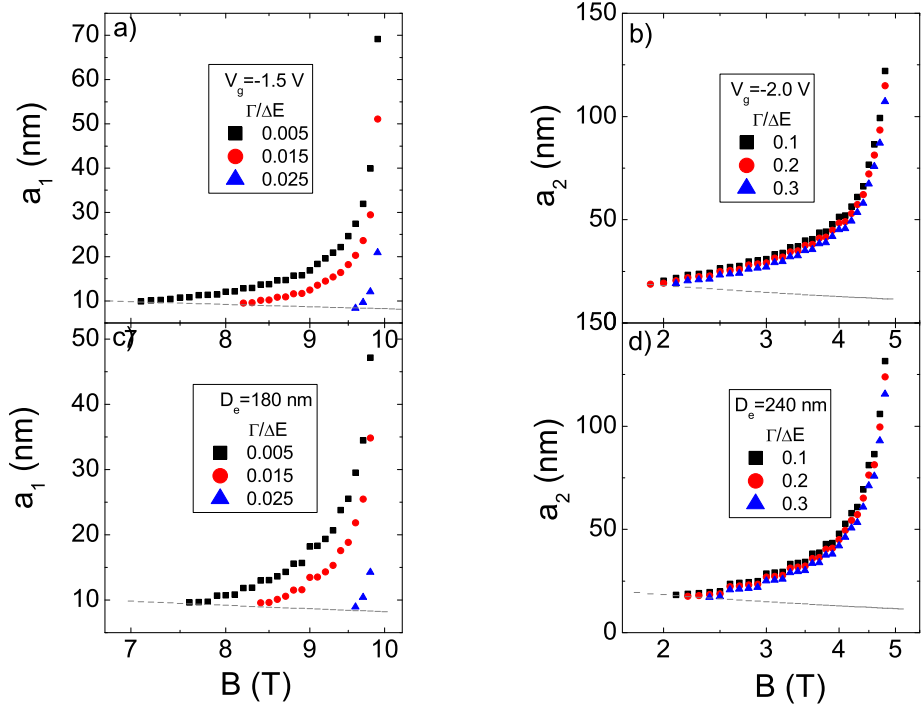}
\caption{(Color online) The evolution of incompressible strips also taking account the effect of collision broadening $\Gamma$
both considering the gate and the etched defined samples.}\label{disorder}
\end{center}
\end{figure}
In a final investigation, we calculate the magnetic field dependency
of the incompressible strips also considering the effects due to
collision broadening, Fig.~\ref{disorder}. We included the effect of
disorder via density of states broadening, however, the effect of
long-range potential fluctuations stemming from impurities is
neglected. This can be justified, if one considers a high-mobility
system. Whereas,
 the low-mobility samples can also be modeled within the screening theory once the long-range fluctuations are included to
 the self-consistent scheme~\cite{Sinem}.
 We see that incompressible
strips do not exist in all magnetic fields, since the gap collapses
due to scattering broadening. The critical magnetic field value that
the incompressible strip collapses is determined by the electron
density, gate voltage, etching depth etc. The odd incompressible
strips widths are narrower compared to even incompressible strips
widths and it stems from the fact that the Zeeman gap is small
compared to the Landau gap, if exchange effects are neglected.  The
$\nu=1$ incompressible strip forms in magnetic field interval $7-10$
T, whereas $\nu=2$ incompressible strip is formed in magnetic field
interval $2-5$ T. We see that stronger disorder results in a
narrower incompressible strip. At $\Gamma=0.1$ $\hbar\omega_c$,
$0.2$ $\hbar\omega_c$, $0.3$ $\hbar\omega_c$ disorder energies the
$\nu=1$ incompressible strip cannot form, whereas while $\nu=2$
incompressible strip is present. It is clear that the effect of
disorder is much more crucial for the odd integer filling factors.
Numerically, we find that to observe an incompressible strip
assuming $\nu=1$ the disorder potential strength must be lower than
$0.025$ $\hbar\omega_c$, whereas for $\nu=2$ this value is changed
to $0.3$ $\hbar\omega_c$. Basically, disorder broadens the density
of states and when disorder potential leads the wave functions to
overlap on both sides of the incompressible strip, essentially the
incompressible strip vanishes.

\section{Conclusions}
\label{conc}We investigated the details of Chklovskii \emph{et
al}~\cite{Chklovskii92:4026} picture by
taking into account the full electrostatics of the samples. By
considering the device properties, \emph{e.g.} geometry, disorder,
confinement, boundary conditions also taking into account many body and disorder effects, qualitatively, it is shown that the estimations of the non-self-consistent picture is strongly limited. Instead, by performing self-consistent calculations, we
derived semi-empirical formulas for realistic electron density
distributions utilizing a 3D numerical algorithm. In
addition, the variation of the depletion lengths is studied
considering etched and gate defined samples. We observed that, the
linear gate voltage-depletion length dependency assumed in
Ref.~\cite{Chklovskii92:4026} is not justified. It is also found
that the etch and the gate defined samples present fairly different
profiles compared to existing literature. We found that the
electron density presents a steep distribution in etched samples. Whereas, gated samples
provide a full control of edge profile together with a smooth electron
density distribution.
As a final remark, the non-self-consistent solution leads very wide incompressible strips,
hence, while calculating the widths of the incompressible strips,
one should carefully take into account the effects stem from self-consistent calculations to be able to describe the edges of the system under experimental investigation.

\bibliographystyle{elsarticle-num}

\newpage
\begin{center}
\textbf{FIGURE CAPTIONS}
\end{center}
\vspace{1.0cm}

\noindent \textbf{Figure 1:} (a) Schematic presentation of the
crystal. (b) The curves depict the confinement energy (thick black
curve), the Landau levels (broken lines following the confinement
potential) and the Fermi energy (horizontal dashed line),
respectively. Red thin parabolic curve and red thick lines depict
the magnetic confinement and the position dependent Landau levels,
respectively.
(c) The estimated plot of filling factor as a function of position from (b).\\

\noindent \textbf{Figure 2:} (Color online) Electron density
distribution of a realistic sample as a function of position up to
the bulk, starting from the physical boundary. (a) The gate defined
sample, (b) the etch defined sample and (c) the trench-gated sample.
The heterostructure parameters, such as the depth of
the electron layer, are fixed whereas the gate potential and etching depth are varied. \\

\noindent \textbf{Figure 3:} (Color online) Spatial distribution of
the electron density with different heterostructure parameters.
Solid line corresponds to gated samples and dashed line etched
samples. A negative voltage -2.0 V is applied to the gates, whereas
for the etched samples, the depth of etching is fixed to 150 nm. The
distances between electron gas and surface $d_{2DEG-Srf}$ are taken
to be 151 nm, 216 nm, 288 nm. The corresponding depths of the donor
layers are $86.4$ nm, $122.4$ nm and nm, $165.6$ nm, respectively.
Similar calculations are also performed considering $d_{2DEG-Srf}=$
180nm and 266 nm, where
$d_{donor-Srf}=$ 100.8 nm and 151.2 nm, however, the tendencies are left unchanged.\\

\noindent \textbf{Figure 4:} Comparison of densities which are
obtained by analytical formulation, self-consistent calculation
(SCC) and Chklovskii
formalism (CSG).\\

\noindent \textbf{Figure 5:} (a) Schematic representation of the
formation of incompressible strip $\nu=1$ at $x_1$. (b) The
variation of the total potential with respect to spatial
coordinates. The solid thick line and shifted thin lines represent
background potential and Landau levels, respectively. Grey region
indicates the enlargement of Landau Levels
due to disorder, by an amount of $\Gamma$, whereas the dashed horizontal line is Fermi energy.\\

\noindent \textbf{Figure 6:} Incompressible strip widths of $\nu=1$
and $\nu=2$ considering different Lande-$g^*$ factors. The strip
widths with black lines correspond to the Chklovskii formalism and
red lines correspond to the self-consistent
calculation.\\

\noindent \textbf{Figure 7:} (Color online) The evolution of
incompressible strips also taking account the effect of collision
broadening $\Gamma$ both considering the gate and the etched defined
samples.\\

\end{document}